# Representative Functional Connectivity Learning for Multiple Clinical groups in Alzheimer's Disease

Lu Zhang[1], Xiaowei Yu[1], Yanjun Lyu[1], Li Wang[2], Dajiang Zhu[1]

[1]Computer Science and Engineering, University of Texas at Arlington, Arlington, TX, USA
`lu.zhang2@mavs.uta.edu`
[2]Mathematics, University of Texas at Arlington, Arlington, TX, USA

**Abstract.** Mild cognitive impairment (MCI) is a high-risk dementia condition which progresses to probable Alzheimer's disease (AD) at approximately 10% to 15% per year. Characterization of group-level differences between two subtypes of MCI – stable MCI (sMCI) and progressive MCI (pMCI) is the key step to understand the mechanisms of MCI progression and enable possible delay of transition from MCI to AD. Functional connectivity (FC) is considered as a promising way to study MCI progression since which may show alterations even in preclinical stages and provide substrates for AD progression. However, the representative FC patterns during AD development for different clinical groups, especially for sMCI and pMCI, have been understudied. In this work, we integrated autoencoder and multi-class classification into a single deep model and successfully learned a set of clinical group related feature vectors. Specifically, we trained two non-linear mappings which realized the mutual transformations between original FC space and the feature space. By mapping the learned clinical group related feature vectors to the original FC space, representative FCs were constructed for each group. Moreover, based on these feature vectors, our model achieves a high classification accuracy – 68% for multi-class classification (NC vs SMC vs sMCI vs pMCI vs AD).

## 1 Introduction

Alzheimer's disease (AD) is one of the top 10 causes of death in America. After established, AD cannot be cured or even slowed [1]. Mild cognitive impairment (MCI), as a prodromal stage of AD, is a high-risk dementia condition which progresses to probable AD at approximately 10% to 15% per year [2]. Among all the MCI patients, someone will finally convert to probable AD while others will not. Accordingly, MCI is divided into two subtypes – stable MCI (sMCI) and progressive MCI (pMCI) [3]. Characterization of group-level differences between sMCI and pMCI subjects is the key step to understand the underlying mechanisms of MCI progression and enable possible delay of transition from MCI to AD. Among different imaging based measurements, functional connectivity (FC) is considered as a promising way to study MCI progression, since which may show alterations even in preclinical stages [4] and provide substrates for AD progression [5]. For example, FC is widely used to explore MCI related network level alternations and some meaningful biomarkers have been found in MCI patients [6-8]. However, the representative FC patterns for different clinical groups during AD progression, especially for sMCI and pMCI, have been understudied. It worth noting that simply averaging the FCs can be limited in capturing the group-specific patterns because of individual variability and the smoothing effects.

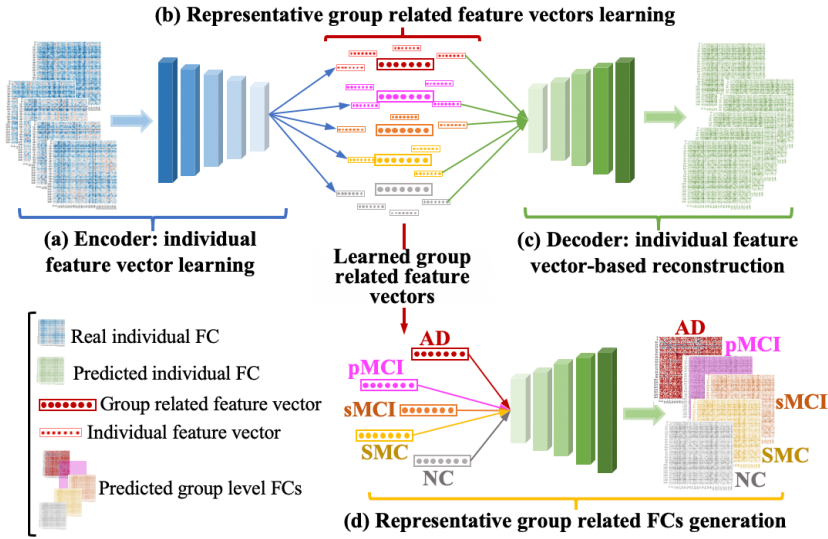

**Fig. 1. (a):** The encoder model learns a non-linear mapping to transform the individual FCs to the feature space and get the individual feature vectors. **(b):** By integrating a distance-based multi-class classification task into autoencoder model, a set of representative group related feature vectors are gradually learned. **(c):** Taking the individual feature vectors and the corresponding original FCs as inputs, the decoder model is trained to learn a non-linear mapping from feature space to the original FC space via a regression task. **(d):** After the model is well-trained, group level feature vectors are fed into the decoder model to generate representative group related FCs.

In this work, we aimed to develop an effective framework to generate representative functional connectivity for AD related clinical groups, including normal control (NC), significant memory concern (SMC), stable MCI (sMCI), progressive MCI (pMCI) and AD. Figure 1 illustrates the core idea of our approach. Firstly, we trained a deep encoder network to extract meaningful features buried in the intrinsic complex and non-linear relations in FC matrix. As a result, the encoder model learned a non-linear mapping from original FC space to feature space. To capture the intrinsic patterns of each clinical group, we proposed a distance-based loss function and integrated a multi-class classification task into the feature extracting process in the feature space. By jointly training, disease related individual feature vectors and representative feature vectors of each group were generated. Then, we introduced a regression task into the model to learn a non-linear mapping from feature space to the original FC space. In general, we integrated autoencoder and multi-class classification into a single deep model and successfully learned a set of representative group related feature vectors and two non-linear mappings which realized the mutual transform between original FC space and the feature space. We applied the well-trained decoder model to the learned group related feature vectors to construct the representative FC for each group. Moreover, based on these feature vectors, our model achieves a high classification accuracy – 68% for multi-class classification (NC vs SMC vs sMCI vs pMCI vs AD).

## 2    Method

### 2.1    Participants and Data Preprocessing

We used 203 subjects (50AD / 27pMCI / 42sMCI / 34SMC / 50NC) from ADNI dataset (http://adni.loni.usc.edu/). Each subject has both T1 and rs-fMRI data. The sMCI and pMCI were selected according to the criteria in [9]. The preprocessing pipeline is as follows: for T1 image, we applied skull removal, segmentation via FreeSurfer (https://surfer.nmr.mgh.harvard.edu/) package and ROI labeling by Destrieux Atlas. The whole brain cortex is divided into 148 regions. For rs-fMRI image, we applied spatial smoothing, slice time correction, temporal pre-whitening, global drift removal and band pass filtering (0.01-0.1 Hz) using FEAT command in FMRIB Software Library (FSL) (https://fsl.fmrib.ox.ac.uk/fsl/fslwiki/). After the preprocessing, we applied Pearson Correlation Coefficient to regional averaged fMRI signals and created FC matrices for each subject.

### 2.2    Method Overview

We proposed a deep encoder-decoder model to learn the representative group related functional connectivity (FC) for each AD related clinical group. There are two major steps in the proposed method: 1) we built an encoder model to transform individual FCs to feature space to generate individual feature vectors. To learn representative features for each clinical group, we parameterized a feature vector for each group in the feature space. By fusing a distance-based multi-class classification task with autoencoder in feature space, intrinsic patterns of different groups are captured by both of the individual feature vectors and group related feature vectors (*Section 2.3*); 2) We trained a decoder model to learn a non-linear mapping from feature space to original FC space. We applied the learned non-linear mapping to the learned group related feature vectors and generated the representative group related FCs (*Section 2.4*).

### 2.3    Representative Group related Feature vector learning

Given the input data $X = \{x_i \in R^d\}_{i=1}^n$, the encoder model transforms $X$ to a feature space via a nonlinear mapping $f(x_i, W_e): R^d \to R^k$ with model parameter $W_e$. To main-tain representative features in feature space for different clinical groups, we parameterized and initialized feature vectors for each group which can be represented by $Z = \{z_c \in R^k\}_{c=1}^{C}$, where $C$ is the number of clinical groups. To connect the feature vectors with the intrinsic patterns of each clinical group, a distance-based classification task is integrated into the autoencoder model.

Specifically, given an input $x_i$ and the corresponding label $y_i$, firstly, we transformed $x_i$ to the feature space and obtained its feature vector $f(x_i, W_e)$. Then, we calculated the distance between $f(x_i, W_e)$ with each one of the $C$ group related feature vectors $z_c$ and assigned $x_i$ to the group whose feature vector is closest to $x_i$. The predicted group label $y_i'$ can be formulated as:

$$y_i' = \underset{c \in \{1,2,\cdots,C\}}{\mathrm{argmin}} \|f(x_i, W_e) - z_c\|^2 \quad (1)$$

In our deep modeling, each label is transformed into a row vector – $\vec{y_i'}$, which is represented in one-hot manner with the $c^{th}$ item of the vector for the appearance of the $c^{th}$ class, that is $\vec{y_{i,c}'} = 1$ if $y_i' = c$ otherwise $\vec{y_{i,c}'} = 0$. Based on the input data and model parameters, we proposed a distance-based classification loss:

$$L_c(\{x_i, y_i\}; f, Z) = -\frac{1}{n} \sum_i \sum_c \vec{y_{i,c}'} \log \frac{\exp\{-\gamma \|f(x_i, W_e) - z_c\|^2\}}{\sum_{j=1}^{C} \exp\{-\gamma \|f(x_i, W_e) - z_j\|^2\}} \quad (2)$$

where $\gamma$ is a regularization parameter to control the hardness of the distance. From (1) and (2), we can see that minimizing the distance-based classification loss essentially means decreasing the distance between the feature vector $f(x_i, W_e)$ and the group related feature vector $z_c$ which is from the same group with $x_i$. Guided by this objective, encoder model parameters $W_e$ and group related feature vectors $z_c$ will be jointly trained and the clinical group related intrinsic patterns are gradually captured by both individual feature vectors and group related feature vectors.

### 2.4 Representative Group related FCs generation

With representative group related feature vectors in hand, to generate representative FC for each group, we need a mapping from feature space to the original FC space. To achieve this goal, we trained a decoder model to learn a non-linear mapping from feature space to the original FC space based on the learned disease related individual feature vectors and the corresponding individual FCs. Then we applied the learned decoder model to the group related feature vectors to generate representative group related FC.

Specifically, the decoder model can be formulated by another non-linear function $g(x_i, W_d): R^k \to R^d$, which maps $f(x_i, W_e)$ back to its original space. In general, the decoder is a mirrored version of the encoder. The loss function of the decoder model is formulated by minimizing the reconstruction error with respect to $\{f, g\}$ given by:

$$L_r = \frac{1}{n} \left( \sum_i \|x_i - g(f(x_i, W_e), W_d)\|^2 + \delta L_{pcc}(x_i, g(f(x_i, W_e), W_d)) \right) \quad (3)$$

Inspired by previous work [10] that both the magnitude and the structure are important to generate a high-quality connectivity matrix, we used the same $L_{pcc}$ loss function as in [10] as a regularization term in our work. The hyper-parameter $\delta$ is introduced to trade-off the power of the magnitude constraint and the structure constraint. The whole model including the encoder part, classification part and the decoder part was trained together in an end-to-end manner. After the model was well trained, we applied the decoder model to the group related feature vector – $z_c$, and get the representative FC – $g(z_c, W_d)$ for each group.

## 3  Results

We applied our proposed deep encoder-decoder model to individual FCs and conducted five-fold cross validation in this work. To improve the accuracy and reduce the influence of random noise, we repeated the experiments 100 times. The experimental setting is shown in Section 3.1. Based on the 100 sets of results, we calculated the representative FC matrix for each group. We analyzed the distribution of the values contained in the five representative matrices and found the top connectivity with the greatest variability across groups. We also displayed the top changed connectivity with the highest frequency in the repeated 100 experiments. All these results are shown in Section 3.2. Both distance-based classification (*Section 2.3*) and the individual FC reconstruction (*Section 2.4*) play important roles during the group level FC learning process, since the former determines whether the intrinsic patterns of each group is effectively captured while the latter governs the quality of the FC construction. At last, we evaluated the classification and regression performance in Section 3.3 and 3.4, respectively.

### 3.1  Experimental Setting

The encoder model is a fully connected multilayer perceptron with dimensions 512-512-256-64-32-20. The decoder is a mirrored version of the encoder. Two hyper-parameters $\gamma$ and $\delta$ are set to be 1. The parameters of the model are initialized by the Xavier scheme. Activation function *Relu* and *batchnorm* were applied at each layer. The Adam optimizer was used to train the whole model with standard learning rate 0.001, weight decay 0.01, and momentum rates (0.9, 0.999).

### 3.2  Representative FC of Each Clinical Group

For each experiment, we obtained five FCs corresponding to five clinical groups and by repeating the experiments 100 times, we got 100 sets of FCs in total. The representative FC for each group was calculated by the averaged FC over the 100 results and shown in Fig. 2 (a). To better demonstrate the variabilities across different clinical groups, we extracted and enlarged one patch at the same location of the five FC matrices. As shown in Fig. 2 (a), from NC to AD, the matrix becomes "red" gradually which means positive correlation in the FC matrix is decreasing while the negative correlation is increasing. To verify this, we used 0.02 as interval to display the distribution of the values for the five representative FCs and showed the results in Fig. 2 (b). It is obvious that the proportion of positive values of NC and SMC groups are higher than that of sMCI, pMCI and AD groups. To characterize the differences between sMCI and pMCI groups, we further calculated the distribution of the two groups at a smaller interval – 0.01 and showed the results in Fig. 2 (c). Compared to pMCI, sMCI has more positive values which represents more positive correlation in the sMCI FC matrix. Based on current results, we aimed to find the top changed connectivity with greatest variability across different groups by two strategies. Firstly, we calculated the standard deviation of the five representative FCs and displayed the top 15 changed connectivity in Fig. 2 (d). Secondly, we calculated the standard deviation of the five FCs generated in each of the 100 experiments and selected the top 100 changed connectivity. Then, from these top 100 changed connectivity, we selected the high-frequency ones which appears more than 20 times and showed the results in Fig. 2 (e). The top changed connectivity in Fig.2 (d) and (e) are consistent and most of the regions involved in the top changed connectivity are reported in previous studies for the close relationship to AD, such as the regions in frontal lobe and temporal lobe.

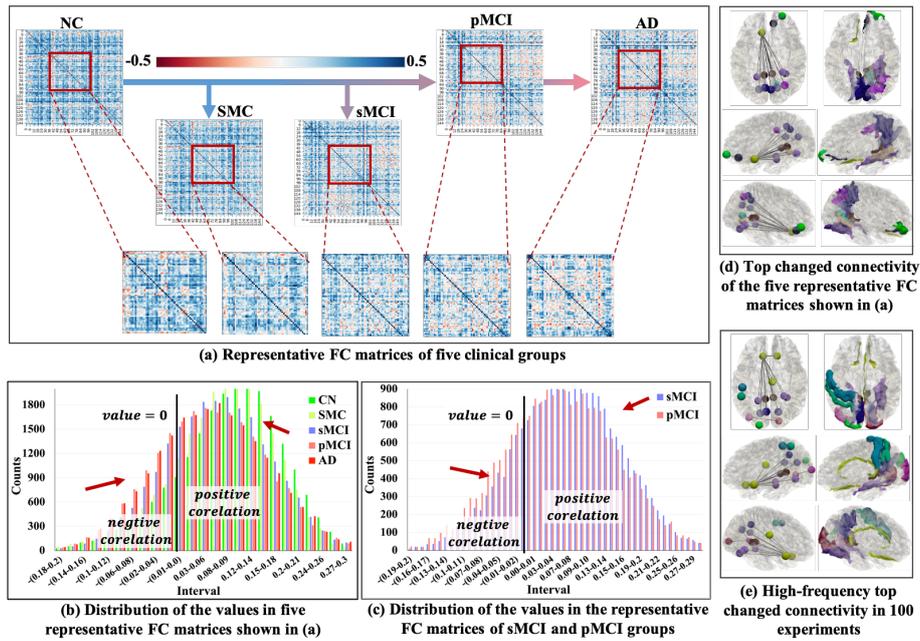

(a) Representative FC matrices of five clinical groups

(b) Distribution of the values in five representative FC matrices shown in (a)

(c) Distribution of the values in the representative FC matrices of sMCI and pMCI groups

(d) Top changed connectivity of the five representative FC matrices shown in (a)

(e) High-frequency top changed connectivity in 100 experiments

**Fig. 2. (a):** We repeated the experiments 100 times and calculated the representative FC for each group. One patch from the same location of each matrix was extracted and the enlarged ones are showed in the bottom. **(b):** Distributions of the values for the five representative FCs. **(c):** Distribution of the values for the representative FCs of sMCI and pMCI group. **(d):** Top 15 changed connectivity across the five representative FCs. The first row shows the brain regions involved in the top changed connectivity. The second row shows the connectivity and the corresponding regions are represented by bubbles with the same color. The colors used in this figure are the same in Destrieux atlas in FreeSurfer. **(e):** top changed connectivity which appear more than 20 times in the 100 experiments.

### 3.3 Classification Performance

We used two strategies to evaluate the proposed method. Firstly, we used the same dataset and five-fold cross validation to conduct experiments with four broadly used machine learning methods including *support vector machine (SVM)*, *k-nearest neighbors (KNN)*, *logistic regression* and *random forest*. The classification performance was measured through calculating class-specific $F_1$ scores and accuracy (Acc). The results are showed in Table 1 (A). We can see that the class-specific $F_1$ scores of our model is over 0.60 and for some classes it can reach 0.70, which is outstanding in multi-class classification of MCI progression and significantly outperforms the other four methods. Secondly, we compared the classification performance with three latest deep learning methods on MCI progression and reported the results in Table 1 (B). [11] obtains good accuracy for AD group and [13] shows good accuracy for NC group. However, both of them only achieve good performance on some specific groups. But for most of the other groups, the accuracy is low. Our methods achieve good performance for all group, which indicates that the learned individual feature vectors and the representative group related feature vectors capture the intrinsic patterns of corresponding clinical stages.

**Table 1.** AD related Multi-Class Classification Performance Comparison with other Methods.

| (A) Comparison with traditional machine learning methods | | | | | | | |
|---|---|---|---|---|---|---|---|
| Method | F1 (all) | F1 (AD) | F1 (pMCI) | F1 (sMCI) | F1 (SMC) | F1 (NC) | Acc (all) |
| SVM | 0.531 ±0.03 | 0.541 ±0.06 | 0.505 ±0.05 | 0.543 ±0.06 | 0.564 ±0.05 | 0.523 ±0.04 | 0.508 ±0.03 |
| KNN | 0.569 ±0.05 | 0.438 ±0.09 | 0.484 ±0.05 | 0.563 ±0.07 | 0.548 ±0.08 | 0.554 ±0.04 | 0.582 ±0.04 |
| Logistic Regression | 0.539 ±0.03 | 0.538 ±0.06 | 0.458 ±0.04 | 0.534 ±0.05 | 0.538 ±0.05 | 0.553 ±0.03 | 0.535 ±0.02 |
| Random Forest | 0.462 ±0.03 | 0.364 ±0.05 | 0.536 ±0.07 | 0.468 ±0.04 | 0.435 ±0.06 | 0.346 ±0.04 | 0.424 ±0.03 |
| **Proposed Model** | **0.674** ±0.01 | **0.716** ±0.03 | **0.639** ±0.03 | **0.637** ±0.06 | **0.681** ±0.02 | **0.695** ±0.03 | **0.676** ±0.02 |
| (B) Comparison with deep learning methods | | | | | | | |
| Work | Modality | | Participants | | | Performance (Acc) | |

| | | | |
|---|---|---|---|
| Liu et al. (2014) [11] | MRI, PET | 77NC, 102sMCI, 67pMCI, 85AD | AD: **0.64**; pMCI: 0.53; sMCI: 0.52; NC: 0.59 |
| Shi et al. (2017) [12] | MRI, PET | 51AD, 56sMCI, 43pMCI, 52NC | Total: 0.57 (NC/sMCI/pMCI/AD) |
| Zhou et al. (2019) [13] | MRI, PET, SNP | 190AD, 226HC, 157pMCI, 205sMCI | AD: 0.57; pMCI: 0.62; sMCI: 0.34; NC: **0.63** |
| **Proposed Model** | rs-fMRI | 50AD, 27pMCI, 42sMCL, 34SMC, 50NC | AD: **0.71**; pMCI: **0.69**; sMCI: **0.65**; SMC: **0.67**; NC: **0.66**; Total: **0.68** |

## 3.4 Regression Performance

To evaluate the regression performance of the proposed model, we calculated two mean MSE for each subject – mean MSE between real individual FC with group population averaged FC and mean MSE between real individual FC with predicted individual FC generated by our model. We calculated the difference between the two MSE values of each subject and showed the results in Fig. 3 left part by line chart. To better display the results, the positive values which indicate the predicted FCs have smaller mean MSE were represented by blue lines and the negative values which indicate the averaged FC have smaller mean MSE were represented in orange lines. From the results we can see that most of the predicted FCs generated by our model have smaller MSE. We demonstrated the distributions of the two sets of mean MSE by violin plot and performed significance analysis with p-value. The results were showed in Fig. 3 right part. The individual FC predicted by our proposed model is significantly ($p < 0.0002$) better than the group population averaged FC.

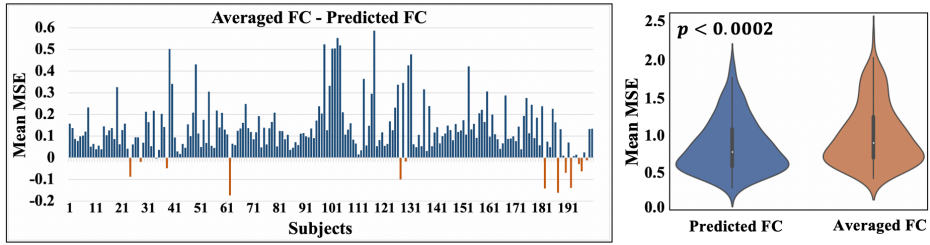

**Fig. 3.** Comparison of the mean MSE between the predicted individual FC and group averaged FC. **Left**: For each subject, we calculated two mean MSE – mean MSE between real individual FC with group averaged FC and mean MSE between real individual FC with predicted individual FC by our model. We should the difference of the two MSE values of each subject on the left part. **Right**: The distributions of the two sets of mean MSE were demonstrated by violin plot at the right side and the p-values were also calculated.

## 4  Conclusion and Discussion

In this work, we proposed a novel deep encoder-decoder framework to generate representative functional connectivity matrix for each AD related clinical group. we integrated autoencoder and multi-class classification into a single deep model and learned a set of representative group related feature vectors as well as two non-linear mappings to realize the mutual transform between original FC space and the feature space. By applying the well-trained decoder model to the learned group related feature vectors, representative FCs were generated for each AD related clinical groups. Moreover, based on the learned feature vectors, our proposed model achieves a high prediction performance over 68% for multiple AD related groups.

**References**

1. Ferri, C.P., Prince, M., Brayne, C., Brodaty, H., Fratiglioni, L., Ganguli, M., Hall, K., Hasegawa, K., Hendrie, H., Huang, Y., et al.: Global prevalence of dementia: a delphi consensus study. The lancet 366(9503), 2112–2117 (2005)
2. Petersen, R.C.: Mild cognitive impairment as a diagnostic entity. Journal of internal medicine 256(3), 183–194 (2004)
3. Petersen, R. C., Smith, G. E., Waring, S.C., Ivnik, R. J., Tangalos, E. G., Kokmen, E.: Mild cognitive impairment: clinical characterization and outcome. Archives of neurology 56(3), 303–308 (1999)
4. Mormino, E.C., Smiljic, A., Hayenga, A.O., H. Onami, S., Greicius, M.D., Rabinovici, G.D., Janabi, M., Baker, S.L., V. Yen, I., Madison, C.M., et al.: Relationships between beta-amyloid and functional connectivity in different components of the default mode network in aging. Cerebral cortex 21(10), 2399–2407 (2011)
5. Franzmeier, N., Dyrba, M.: Functional brain network architecture may route progression of alzheimer's disease pathology. Brain 140(12), 3077–3080 (2017)
6. Brier, M.R., Thomas, J.B., Fagan, A.M., Hassenstab, J., Holtzman, D.M., Benzinger, T.L., Morris, J.C., Ances, B.M.: Functional connectivity and graph theory in preclinical alzheimer's disease. Neurobiology of aging 35(4), 757–768 (2014)
7. Wang, K., Liang, M., Wang, L., Tian, L., Zhang, X., Li, K., Jiang, T.: Altered functional connectivity in early alzheimer's disease: A resting-state fmri study. Human brain mapping 28(10), 967–978 (2007)
8. Liu, Y., Yu, C., Zhang, X., Liu, J., Duan, Y., Alexander-Bloch, A.F., Liu, B., Jiang, T., Bullmore, E.: Impaired long distance functional connectivity and weighted network architecture in alzheimer's disease. Cerebral Cortex 24(6), 1422–1435 (2014)
9. Risacher, S.L., Saykin, A.J., Wes, J.D., Shen, L., Firpi, H.A., McDonald, B.C.: Baseline mri predictors of conversion from mci to probable ad in the adni cohort. Current Alzheimer Research 6(4), 347–361 (2009)
10. Zhang, L., Wang, L., Zhu, D.: Recovering brain structural connectivity from functional connectivity via multi-gcn based generative adversarial network. In: International Conference on Medical Image Computing and Computer-Assisted Intervention. pp. 53–61. Springer (2020).
11. Liu, S., Liu, S., Cai, W., Che, H., Pujol, S., Kikinis, R., Feng, D., Fulham, M.J., et al.: Multimodal neuroimaging feature learning for multiclass diagnosis of alzheimer's disease. IEEE Transactions on Biomedical Engineering 62(4), 1132–1140 (2014)

12. Shi, J., Zheng, X., Li, Y., Zhang, Q., Ying, S.: Multimodal neuroimaging feature learning with multimodal stacked deep polynomial networks for diagnosis of alzheimer's disease. IEEE journal of biomedical and health informatics 22(1), 173–183 (2017)
13. Zhou, T., Thung, K.H., Zhu, X., Shen, D.: Effective feature learning and fusion of multimodality data using stage-wise deep neural network for dementia diagnosis. Human brain mapping 40(3), 1001–1016 (2019)